%% file: MICCAI2025_paper_template.tex
\documentclass[runningheads]{llncs}
\usepackage{amsmath}
\usepackage{amssymb}
\usepackage{multirow}
\usepackage{graphicx}
\usepackage{booktabs} 
\usepackage{adjustbox}
\usepackage{amsmath}  
\usepackage{xcolor}  
\usepackage{marvosym}
\usepackage[T1]{fontenc}

\usepackage{graphicx,verbatim,amsmath}

\usepackage{color}
\usepackage[colorlinks,linkcolor=blue,citecolor=blue,urlcolor=blue]{hyperref}

\begin{document}

\title{Instance Migration Diffusion for Nuclear Instance
Segmentation in Pathology}

\author{Lirui Qi, Hongliang He\textsuperscript{(\Letter)}, Tong Wang, Siwei Feng, Guohong Fu} 
\authorrunning{Qi et al.}
\institute{Schools of Computer Science and Technology, Soochow University, Suzhou, China\\
    \email{hlhe2023@suda.edu.cn}}

\maketitle              

\begin{abstract}
Nuclear instance segmentation plays a vital role in disease diagnosis within digital pathology. However, limited labeled data in pathological images restricts the overall performance of nuclear instance segmentation. To tackle this challenge, we propose a novel data augmentation framework Instance Migration Diffusion Model (IM-Diffusion), IM-Diffusion designed to generate more varied pathological images by constructing diverse nuclear layouts and internuclear spatial relationships. In detail, we introduce a Nuclear Migration Module (NMM) which constructs diverse nuclear layouts by simulating the process of nuclear migration. Building on this, we further present an Internuclear-regions Inpainting Module (IIM) to generate diverse internuclear spatial relationships by structure-aware inpainting. On the basis of the above, IM-Diffusion generates more diverse pathological images with different layouts and internuclear spatial relationships, thereby facilitating downstream tasks. Evaluation on the CoNSeP and GLySAC datasets demonstrate that the images generated by IM-Diffusion effectively enhance overall instance segmentation performance. Code will be made public later.


\end{abstract}

\section{Introduction}
Pathological images are widely regarded as the gold standard in diagnosis, where nuclei segmentation and classification serve as critical tasks. Recent advances in deep learning have established model-based approaches as mainstream solutions for these tasks \cite{unet,cdnet,Hover}. For instance, HoVer-Net \cite{Hover} enhances segmentation accuracy by leveraging distance information, while the Prompted Nuclei Segmentation (PNS) \cite{shui2025unleashing}, based on Segment Anything Model (SAM) \cite{kirillov2023segment} demonstrates promising adaptability. However, the high annotation cost of histopathological images restricts the availability of annotated training data. This limitation hinders the performance of deep learning models in segmentation tasks \cite{cong2024adaptive}.

Generative models like Generative Adversarial Networks (GANs) \cite{GAN} have become popular for synthesizing diverse data, offering a data-driven solution to address limited annotations \cite{Lou_2023}. However, GAN-based approaches are prone to mode collapse and training instability \cite{GAN,Lou_2023,abousamra2023topology}, compromising reliability. Denoising diffusion probabilistic models (DDPMs) \cite{ho2020denoising} recently emerged as stable and high-quality alternatives to overcome these limitations. Semantic diffusion model (SDM) \cite{wang2022semantic} enhances controllability through conditional semantic guidance. In histopathology generation, recent works explore specialized applications: DiffMix \cite{oh2023diffmix} addresses nuclear imbalance with Balancing and Enlarging maps, while Nudiff \cite{yu2023diffusion} enables unconditional nuclear synthesis. Extensions like NASDM \cite{shrivastava2023nasdm} integrate multi-conditions generation (instance/semantic), and Oh et al. \cite{oh2024controllable} incorporate text for instance-specific refinement. Despite progress, challenges remain in generating histopathological images that achieve both layout and morphological diversity. In pathological images, the position, layout, and spatial relationships of nuclei are critical factors influencing segmentation and classification \cite{derakhshan2022pathogenesis,abousamra2023topology}. However, relying solely on the original labels and layout limits generative models' ability to capture nuclear features \cite{abousamra2023topology}.


To address these limitations, we propose Instance Migration Diffusion (IM-Diffusion), the framework that enhances the diversity of nuclear layouts and internuclear spatial relationships. Drawing inspiration from empirical observations of tissue manipulation artifacts \cite{zarella2019practical}, we introduce the Nuclear Migration Module (NMM) to simulate nuclear migrations. Consistent with prior observations \cite{Stylianopoulos2011,Ming2015}, smaller nuclei display enhanced mobility under stress, following a power-law relationship characteristic of viscoelastic materials. Building on this foundation, we further identify critical challenges in enhancing the diversity of internuclear spatial relationships. Nuclei in dense internuclear regions exhibit complex morphological characteristics \cite{bell2020reflection,coons2007histopathology}. Conventional augmented methods often overlook internuclear relationships, restricting segmentation accuracy. To mitigate this, we propose Internuclear-regions Inpainting Module (IIM) to preserve layout information while enabling diverse internuclear spatial relationships. Finally, our refinement pipeline generates high-quality image-label pairs for downstream training using the processed labels.

Our contributions are as follows: (1) We propose a generative data augmentation framework IM-Diffusion that synthesizes diverse pathological images. (2) We formulate a size-dependent mobility module NNM to simulate nuclear migrations, thereby generating morphologically diverse layouts. (3) We construct an structure-aware inpainting module IIM to establish diverse internuclear spatial relationships.





\begin{figure*}[t]
\includegraphics[width=\textwidth]{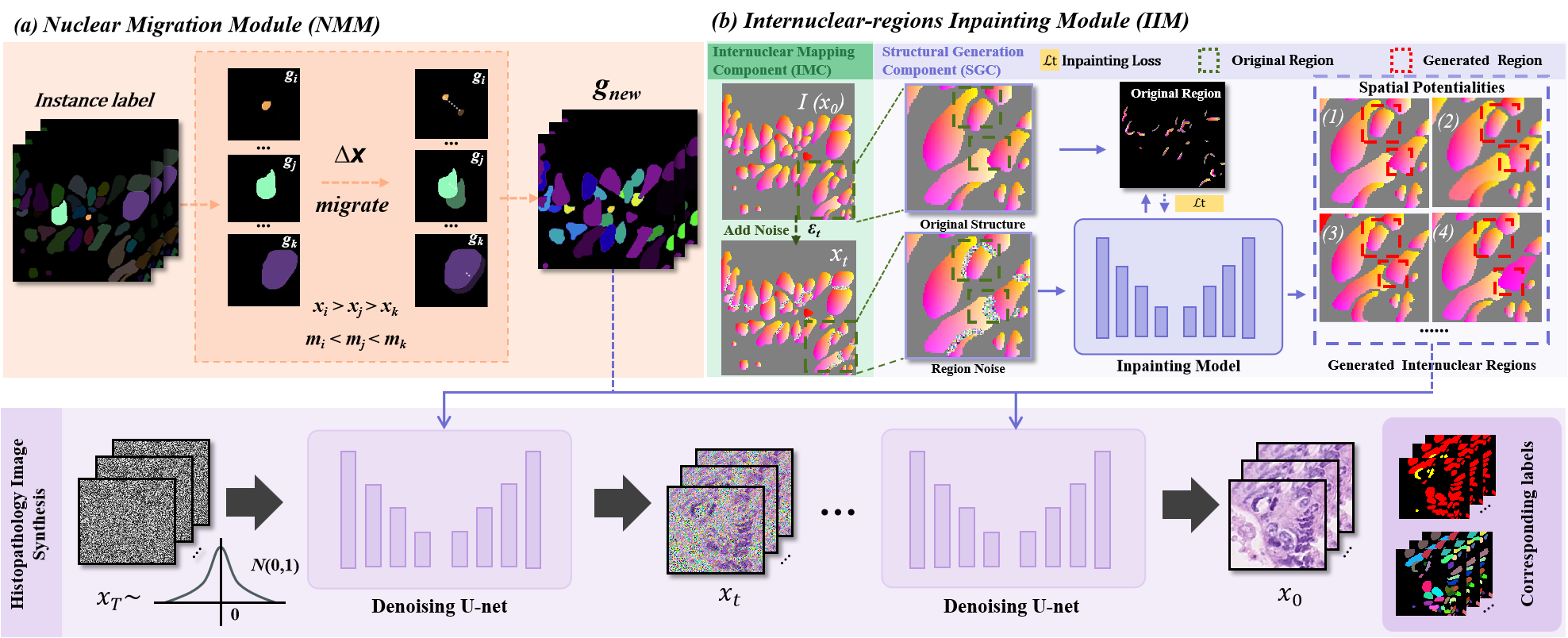}
\caption{The IM-Diffusion framework overview. The IM-Diffusion framework processes instance labels through NMM to generate novel layouts $\mathcal{G}{\text{new}}$. Per-instance $n$ migration distances $ x $ exhibit a negative correlation with nuclear sizes $ m $. Subsequently, the IIM module processes structural labels $ I $ derived from instance labels to create diverse internuclear spatial relationships. This is exemplified by Case 1-2 (overlapping configurations) and Case 3-4 (adhesive nuclear formations). The refined layouts $ \mathcal{G}_{\text{new}} $, processed through IIM, serve as conditioning inputs for image synthesis models, generating image-label pairs that enhance downstream model generalization.
} \label{method_1}
\end{figure*}

\section{Method}
The IM-Diffusion framework, as shown in Figure \ref{method_1}, consists of two key modules: the NMM and the IIM. These modules work together with enhanced the diversity of layouts and internuclear spatial relationships. NMM simulates the nuclear migrations caused by a pathologist manipulating the tissue, while IIM identifies the dense internuclear regions and performs inpainting. After label processing, a pre-trained image diffusion model generates diverse image-label pairs. The following subsections detail module's implementation.


\subsection{Preliminaries}

Diffusion models constitute a probabilistic framework for generative modeling, operating through two fundamental processes \cite{ho2020denoising}: (1) A forward diffusion process that progressively adds Gaussian noise to the data, controlled by a variance parameter $ \beta_t$, and (2) A reverse process that learns to iteratively denoise the data. Specifically, \( \beta_t \) is optimized during training to balance noise injection and signal preservation. The conditional distribution of each state \( x_t \) given the previous state \( x_{t-1} \) is defined as:
\begin{equation}
q(x_t | x_{t-1}) = \mathcal{N}(x_t; \sqrt{1 - \beta_t} \, x_{t-1}, \beta_t I),
\end{equation}

Let $ \alpha_t = 1 - \beta_t $ and $ \bar{\alpha}_t = \prod_{s=1}^t \alpha_s $, representing cumulative variance scaling. The marginal distribution of $ x_t $ given the initial data $ x_0 $ can be written as:
\begin{equation}
q(x_t | x_0) = \mathcal{N}(x_t; \sqrt{\bar{\alpha}_t} \, x_0, (1 - \bar{\alpha}_t) I),
\end{equation}

The \( c_t \) denotes conditioning information for denoising. The function $\mu_\theta$ predicts the mean of the denoising model, and $\sigma_t^2$ is the variance at time step $t$. In the reverse process, a neural network $ \epsilon_\theta $ is employed to estimate noise patterns and recover the original input $ x_0 $ from $ x_t $. The posterior distribution $p_\theta(x_{t-1} | x_t, c_t)$ is expressed as:
\begin{equation}
p_\theta(x_{t-1} | x_t, c_t) = \mathcal{N}(x_{t-1}; \mu_\theta(x_t, t, c_t), \sigma_t^2 I),
\end{equation}


Given \( x_t \) and the corresponding time step \( t \), a trained network can predict the noise distribution derived from the original image. The loss function for network optimization is defined as:

\begin{equation}
\mathcal{L}_{t-1} = \mathbb{E}_{\epsilon \sim \mathcal{N}(0, I)} \left[ \left\| \epsilon - \epsilon_\theta (x_t, y, t) \right\|^2_2 \right]
\end{equation}



\subsection{Nuclear Migration Module}
In the NMM, we simulate nuclear migrations driven by pathologist tissue manipulations. The module first extracts all individual nuclear instances from input instance labels, associating each nucleus $n_i$ with a size parameter $m_i$ (pixel area) and its original instance label $\mathcal{G}_{\text{i}}$. These nuclei form the set $\mathcal{N} = \{n_1, n_2, \ldots, n_K\}$, where $K$ denotes the total number of nuclei in the single label. Empirical studies show smaller nuclei are more mobile under stress, aligning with viscoelastic tissue power-law behavior \cite{zarella2019practical,Stylianopoulos2011}. To mathematically encode this size-dependent behavior, we model the migration magnitude as $\Delta x_i = \Delta X / m_i $ with the shared distance $\Delta X$. The collective motion of all nuclei is characterized through an instance label representation. In this module, each nucleus migrates in the shared random direction $\mathbf{D}$, with the migration distance $\Delta x_i$ being inversely proportional to its size. The final displaced instance label $\mathcal{G}{\text{new}}$ is obtained by superposing the migrated nuclei, as described by the following equation:

\begin{equation}
\mathcal{G}_{\text{new}} = \sum_{i=1}^{K} \frac{\Delta \mathbf{X} \cdot \mathbf{D}}{m_i} \cdot \mathcal{G}_i
\end{equation}

Small nuclei (e.g., inflammatory and 
miscellaneous cells) are particularly critical for accurate segmentation \cite{cong2024adaptive}. Nuclear migration introduces issues that require mitigation: larger nuclei (e.g., stromal cells) may erroneously overlap smaller counterparts (e.g., lymphocytes), completely occluding them. This contradicts histological observations \cite{nam2012density,wu2019acoustofluidic} and restricts the availability of small nuclei for downstream training. Such occlusions hinder further refinement of segmentation accuracy. To address this, we implement dynamic z-ordering based on nuclear size: smaller nuclei receive higher display priority in overlapping regions, ensuring occlusion mitigation while preserving histological fidelity.

\subsection{Internuclear-regions Inpainting Module}
To generate diverse internuclear spatial relationships, our module achieves accurate nuclear structure generation through a inpainting pipeline. The pipeline consists of two main components: (1) The Internuclear Mapping Component (IMC), which generates spatial masks via constrained dilation of nuclear instance labels, and (2) The Structural Generation Component (SGC), which performs structure-aware inpainting using these masks. Specifically, IMC dilates each nuclear label pixel-wise, stopping when it contacts other nuclei and retaining contact pixels to form instance-level internuclear regions $ C_{\text{pre}} $ for each nucleus. Subsequent dilation of $ C_{\text{pre}} $ yields the overall internuclear regions $ C $. SGC captures internuclear spatial relationships via nuclear structure labels $ I $. This labels incorporate two key components: semantic labels \( sem \) and vertical/horizontal distance labels. The distance labels are derived from geometric properties of instance labels. During training, $ I $ serves as input to the inpainting module, while $ C $ acts as a conditioning signal to guide spatial relationship preservation.

For the inpainting framework, we integrate insights from SmartBrush \cite{xie2023smartbrush} and RePaint \cite{lugmayr2022repaint}, proposing a hybrid process that combines forward noise injection and conditional reverse denoising. In the forward noising process, the intermediate image \( x_t \) is not purely random noise. Instead, within the region defined by \( C \), corrupted areas are replaced with the original image perturbed by DDPM noise, while intact regions retain $ x_0 $. Formally, \( x_t \) is defined as follows:



\begin{equation}
x_t = (\sqrt{\bar{\alpha_t}} x_0 + \sqrt{1 - \bar{\alpha_t}}) \odot C   + x_0\odot (1-C)
\end{equation}

Since modifying \( x_t \) significantly alters the denoising signal relative to the original image, training should prioritize the region $ C $ requiring restoration. The optimization objective for the denoising network is formulated as:


\begin{equation}
\mathcal{L}_{t-1} = \mathbb{E}_{\epsilon \sim \mathcal{N}(0, I)} \left[ \left\| \epsilon - \epsilon_\theta (x_t, sem, C, t) \right\|^2_2 \right]
\end{equation}

After denoising \( x_t \) at time step \( t \) to obtain \( x_{t-1} \), the internuclear regions \( C \) are replaced with the original input \( x_0 \), yielding $ x_{t-1}' $ for the next denoising step. This iterative replacement continues until the original image is fully reconstructed. The process is formulated as:



\begin{equation}
x_{t-1}’ = x_{t-1} \odot C   + x_0\odot (1-C)
\end{equation}

\input{table1}

\section{Experiments}
\subsection{Experiments Settings}
\subsubsection{Datasets.}

Experiments are conducted on the CoNSeP \cite{Hover} and GLySAC \cite{doan2022sonnet} datasets. The CoNSeP dataset contains 41 images at 40x magnification and 1000×1000 pixels. According to the original classification scheme \cite{Hover}, we reclassify the 7 types of nuclei into 4 classes. Of these, 27 images are used for training and 14 for testing. The GLySAC dataset contains 59 images at 40x magnification and 1000×1000 pixels, comprising 10 nuclear classes. We reclassify into 3 categories \cite{doan2022sonnet}. For GLySAC, 34 images are used for training and 25 for testing.




\subsubsection{Implementation Details.} Our implementation is based on PyTorch 1.12.1 with two NVIDIA V100 GPUs. For data preprocessing on CoNSeP and GLySAC, we use a sliding window strategy to generate 256×256 pixel image patches with 164-pixel strides. The NMM module employs shared distance dilation $\Delta X$ between 30 to 100 pixels. The IMC component employs 3×3 convolutional kernels for pixel-wise dilation. The SGC component, based on the SDM architecture, processes 256×256 nuclear structure labels as input. This module undergoes 20,000 training iterations with a batch size of 3 per GPU. For image synthesis, the SDM employs Adam optimization with an initial learning rate of $10^{-4}$ and undergoes 10,000 iterations (batch size of 8). For downstream tasks, HoVer-Net and PNS are applied for joint nuclear segmentation and classification, with comprehensive 5-fold cross-validation. HoVer-Net is trained on 256×256 images (batch size 8), maintaining an initial learning rate of $10^{-4}$ that decays every 75 iterations over 200 epochs. In PNS, we preserve data consistency with HoVer-Net's configuration. Both the Prompter and Segmenter components undergo extended training (500 epochs) using the same learning rate ($10^{-4}$). Notably, all comparative experiments use double the original dataset size, except for the baseline model, which uses the full dataset. We conducte ablation studies to analyze the impact of training data volume on performance.

\subsubsection{Evaluation Metrics.}  To evaluate the effectiveness of generated images on downstream models, we employes metrics for nuclear instance segmentation: the multi-class Aggregated Jaccard Index (mAJI) and Panoptic Quality (mPQ). To mitigate potential biases introduce by zero-division during metric computation, we shift to instance-wise evaluation instead of conventional per-image averaging. This approach ensures reliable global model performance by reducing sensitivity to individual image variations. In ablation studies, we retaine the AJI \cite{kumar2019multi} and PQ \cite{Hover} metrics for consistent comparison. For nuclear classification, we use the F1-score as the primary metric, with overall performance quantified by the multi-class F1-score (mF1). During evaluation, PNS uses 256×256 patches (164-pixel stride) to avoid stitching artifacts, while HoVer-Net maintains the baseline 1000×1000 setup. Notably, PNS classification results are reported via outputs from the Prompter to align with our unified analysis pipeline. 

%



\begin{figure*}[t]
\includegraphics[width=\textwidth]{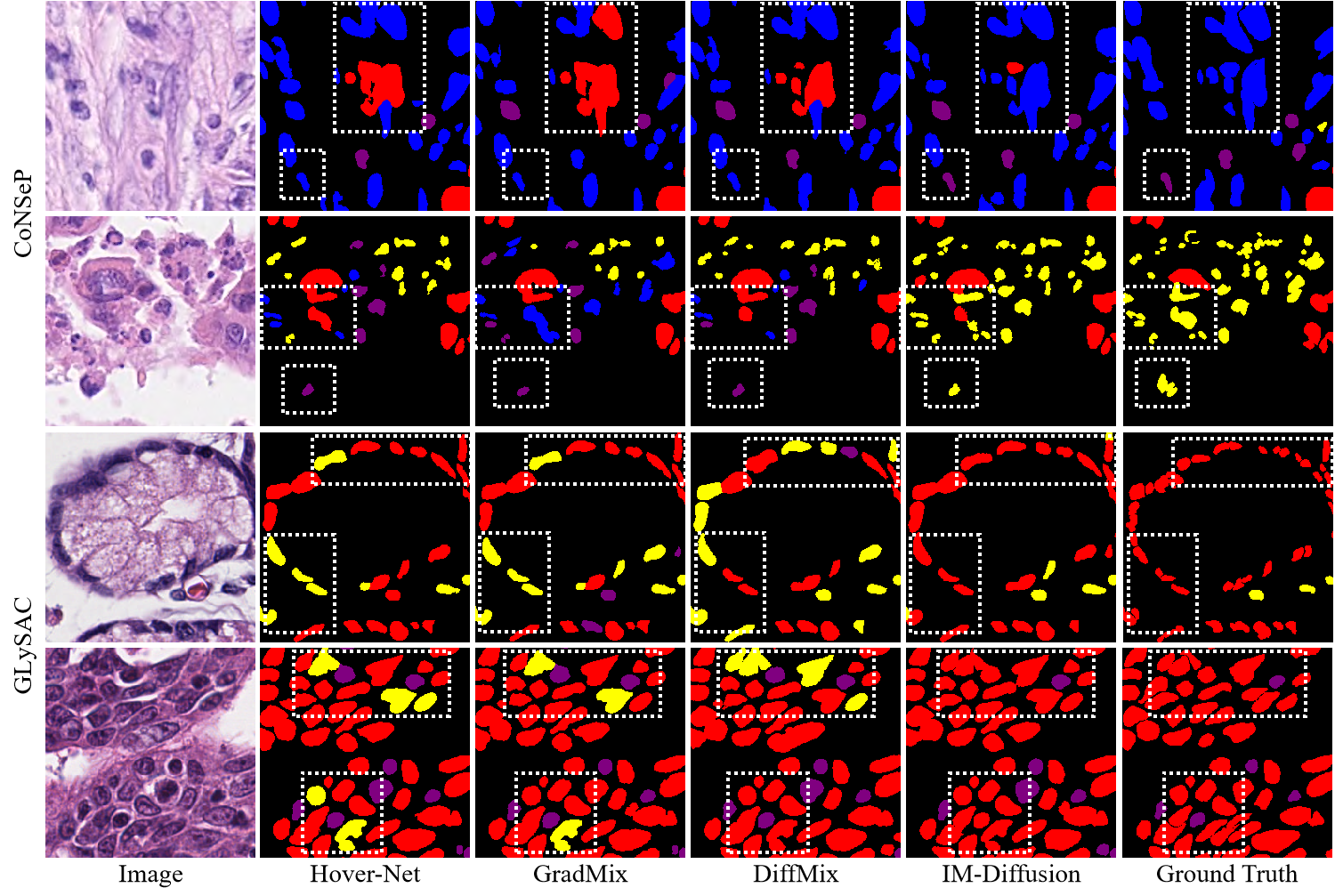}
\caption{Qualitative comparison of instance-level augmentation methods versus IM-Diffusion on histopathology benchmarks. Boxed regions exhibit enhanced classification consistency with Ground Truth (GT) in both CoNSeP and GLySAC datasets.} \label{result_image}
\end{figure*}

\subsection{Results}

The comparative results of IM-Diffusion and existing nuclear instance-level augmentation methods on the CoNSeP and GLySAC datasets are summarized in Table \ref{table1}. Notably, IM-Diffusion achieves state-of-the-art (SOTA) performance, with 48.0\% mAJI and 64.8\% mF1 on CoNSeP, and 43.3\% mAJI and 61.7\% mF1 on GLySAC. A significant improvement is observed in the minority-class classification performance (e.g., F\textsuperscript{m} on CoNSeP), where our method outperforms the baseline by 95.42\%. This demonstrates that providing diverse nuclear layouts and internuclear spatial relationships effectively enhances the model’s understanding of nuclear microenvironments in downstream tasks. 



Visual comparisons of different instance-level augmentation methods are illustrated in Figure \ref{result_image}. As shown, IM-Diffusion improves downstream models' adaptability to nuclear classification tasks, particularly in densely packed internuclear regions. The ablation study results are presented in Table \ref{table2}. Doubling the training data volume yielded no significant performance gains, indicating saturation under the original data distribution. The baseline diffusion model (SDM) provides limited improvements due to its strict adherence to original layout constraints. NMM improves classification metrics by diversifying nuclear layouts, while the IIM module improves segmentation and classification accuracy through diverse internuclear spatial relationships. IM-Diffusion further enhances accuracy by synergistically integrating nuclear layout diversity from NMM with internuclear relationship diversity from IIM.

\input{table2}

\section{Conclusion}

In this study, we propose a novel Instance Migration Diffusion Model (IM-Diffusion) to generate diverse layouts and internuclear spatial relationships in nuclear instance segmentation for digital pathology. Specifically, IM-Diffusion incorporates the Nuclear Migration Module (NMM), which generates diverse nuclear layouts by simulating the nuclear migrations, and the Internuclear-regions Inpainting Module (IIM) generates diverse internuclear spatial relationships through structure-aware inpainting. By enriching pathological images with diverse layouts and internuclear spatial relationships, IM-Diffusion enhances data diversity and expands data distribution, thereby facilitating downstream tasks. Experiments on the CoNSeP and GLySAC datasets validate its effectiveness.

\end{document}

%% file: table1.tex
\begin{table}[t]
\centering
\caption{Experimental results on GLySAC and CoNSeP compared to HoVer-Net and PNS benchmarks. Instance segmentation performance is assessed with class-aware metrics (mAJI/mPQ), while classification accuracy is quantified via multi-class F1-score. Category-specific scores are defined as follows: \( F{\textsuperscript{m}} \) (miscellaneous), \( F{\textsuperscript{i/l}} \) (inflammatory/lymphocyte), \( F{\textsuperscript{e}} \) (epithelial), and \( F{\textsuperscript{s}} \) (spindle). Notably, the spindle category excluded from GLySAC. For CoNSeP, $F^{\text{l/i}}$ represents inflammatory scores, while in GLySAC, $F^{\text{l/i}}$ denotes lymphocytes scores. The best and second-best results are marked in \textcolor{red}{red} and \textcolor{blue}{blue}, respectively.}
\label{table1}
\label{tab:performance_metrics}
\fontsize{8}{10}\selectfont
\begin{tabular}{c c  c c| c c  c c c  }  
\toprule
 & \multirow{2}{*}{\textbf{\parbox[c]{1.2cm}{\centering Method}}} & \multicolumn{2}{c}{\textbf{Segmentation}} & \multicolumn{5}{c}{\textbf{Classification}} \\ \cmidrule{3-9}  
                                      & \textbf{}       & \textbf{mAJI} & \textbf{mPQ}  & \textbf{F\textsuperscript{m}} & \textbf{F{\textsuperscript{i/l}}} & \textbf{F\textsuperscript{e}} & \textbf{F\textsuperscript{s}} & $\textbf{mF1}$ \\ \midrule
\multirow{8}{*}{\rotatebox{90}{CoNSeP}}                                     
& HoVer-Net  & 46.9 $\pm$ 0.2 & 45.7 $\pm$ 0.3  & 23.8 $\pm$ 1.8 & 56.3 $\pm$ 1.3 & 62.1 $\pm$ 1.5 & 53.3 $\pm$ 1.2 & 62.5 $\pm$ 0.3\\ 

& GradMix  & 47.3 $\pm$ 0.5 & 45.8 $\pm$ 0.4  & 28.1 $\pm$ 2.1 & \textcolor{blue}{57.9 $\pm$ 0.9} & \textcolor{blue}{63.2 $\pm$ 1.2} & 53.7 $\pm$ 1.3   & 63.8 $\pm$ 0.2 \\ 
& DiffMix  & \textcolor{blue}{47.7 $\pm$ 0.2} & \textcolor{blue}{46.3 $\pm$ 0.3}  & \textcolor{blue}{29.4 $\pm$ 2.7} & 56.5 $\pm$ 1.5 & \textcolor{red}{63.3 $\pm$ 1.4} & \textcolor{blue}{54.8 $\pm$ 1.3}  & \textcolor{blue}{64.0 $\pm$ 0.3}  \\ 
& IMD     &  \textcolor{red}{48.0 $\pm$ 0.2} & \textcolor{red}{46.7 $\pm$ 0.3} & \textcolor{red}{46.5 $\pm$ 2.2} & \textcolor{red}{58.3 $\pm$ 1.1} & \textcolor{blue}{63.2 $\pm$ 1.2} & \textcolor{red}{54.9 $\pm$ 1.2}   & \textcolor{red}{64.8 $\pm$ 0.2}\\ 

\cline{2-9} 

& PNS       & 48.6 $\pm$ 0.4 & 46.2 $\pm$ 0.3  & 48.5 $\pm$ 2.1 & \textcolor{blue}{76.1 $\pm$ 0.5} & 73.6 $\pm$ 1.3 & 66.3 $\pm$ 0.6 & 66.1 $\pm$ 0.4\\ 
& GradMix   & 48.7 $\pm$ 0.3 & 46.2 $\pm$ 0.3  & 49.1 $\pm$ 1.9 & 76.0 $\pm$ 0.6 & 73.3 $\pm$ 1.3 & 66.4 $\pm$ 0.4 & 66.0 $\pm$ 0.3\\ 
& DiffMix   & \textcolor{blue}{49.2 $\pm$ 0.4} & \textcolor{blue}{46.9 $\pm$ 0.2}  & \textcolor{blue}{49.9 $\pm$ 1.7} & 75.8 $\pm$ 0.6 & \textcolor{blue}{74.2 $\pm$ 1.1} & \textcolor{blue}{66.5 $\pm$ 0.6} & \textcolor{blue}{66.8 $\pm$ 0.5}\\ 
& IMD       & \textcolor{red}{49.4 $\pm$ 0.4} & \textcolor{red}{47.1 $\pm$ 0.3}  & \textcolor{red}{55.6 $\pm$ 1.8} & \textcolor{red}{76.6 $\pm$ 0.4} & \textcolor{red}{74.6 $\pm$ 1.0} & \textcolor{red}{66.8 $\pm$ 0.6} & \textcolor{red}{67.8 $\pm$ 0.5}\\ 
\bottomrule 
\bottomrule 
\multirow{8}{*}{\rotatebox{90}{GLySAC}}                                     
& HoVer-Net & 40.9 $\pm$ 0.2 & 44.7 $\pm$ 0.2  & 27.0 $\pm$ 0.5 & 52.1 $\pm$ 0.3 & 53.23 $\pm$ 0.2 & - & 58.6 $\pm$ 0.2\\
& GradMix    & 40.8 $\pm$ 0.1 & 44.7 $\pm$ 0.1  & 28.5 $\pm$ 0.6 & 51.7 $\pm$ 0.3 & 53.54 $\pm$ 0.1 & - & 58.6 $\pm$ 0.1\\
& DiffMix    & \textcolor{blue}{43.2 $\pm$ 0.2} &\textcolor{red}{ 47.1 $\pm$ 0.2}  & \textcolor{blue}{29.8 $\pm$ 0.4} & \textcolor{red}{54.8 $\pm$ 0.2} & \textcolor{blue}{55.78 $\pm$ 0.3} & - & \textcolor{blue}{60.9 $\pm$ 0.1}\\ 
& IMD        & \textcolor{red}{43.3 $\pm$ 0.2} & \textcolor{blue}{46.7 $\pm$ 0.2}  & \textcolor{red}{32.0 $\pm$ 0.5} & \textcolor{blue}{54.6 $\pm$ 0.3} & \textcolor{red}{57.08 $\pm$ 0.2} & -  & \textcolor{red}{61.7 $\pm$ 0.1}                \\ 

\cline{2-9} 

& PNS       & 43.8 $\pm$ 0.5 & 46.1 $\pm$ 0.5  & 45.9 $\pm$ 0.6 & \textcolor{blue}{68.7 $\pm$ 0.6} & 69.4 $\pm$ 0.4 & - & 61.3 $\pm$ 0.6\\ 
& GradMix   & 43.6 $\pm$ 0.4 & 46.3 $\pm$ 0.3  & 44.4 $\pm$ 1.0 & 67.7 $\pm$ 0.2 & 69.4 $\pm$ 0.4 & - & 60.5 $\pm$ 0.5\\
& DiffMix   & \textcolor{blue}{44.2 $\pm$ 0.4} & \textcolor{blue}{47.1 $\pm$ 0.3}  & \textcolor{blue}{45.9 $\pm$ 0.3} & 68.5 $\pm$ 0.5 & \textcolor{blue}{70.3 $\pm$ 0.4} & - & \textcolor{blue}{61.4 $\pm$ 0.4}\\  
& IMD       & \textcolor{red}{45.2 $\pm$ 0.4} & \textcolor{red}{47.7 $\pm$ 0.4}  & \textcolor{red}{46.4 $\pm$ 0.4} & \textcolor{red}{69.7 $\pm$ 0.4} & \textcolor{red}{72.3 $\pm$ 0.5}    & -      & \textcolor{red}{62.8 $\pm$ 0.5}         \\ 
\bottomrule 
\end{tabular}
\end{table}

%% file: table2.tex
\begin{table}[t]

\centering
\caption{Ablation studies on the CoNSeP. Systematic assessment evaluates training data volume (Marked with a prime (')), base diffusion model, and NMM/IIM module impacts. Apostrophe-marked base models denote doubled training data.}
\label{table2}
\label{tab:performance_metrics}
\fontsize{8}{10}\selectfont
\begin{tabular}{c c  c c c c | c c c c c c}  
\toprule
 & \multirow{2}{*}{\textbf{\parbox[c]{2cm}{\centering Method}}} & \multicolumn{4}{c}{\textbf{Segmentation}} & \multicolumn{6}{c}{\textbf{Classification}} \\ \cmidrule{3-12}  
                                      & \textbf{}       & \textbf{bAJI} & \textbf{mAJI} & \textbf{bPQ} & \textbf{mPQ} & \textbf{F1} & $\textbf{mF1}$ & \textbf{F\textsuperscript{m}} & \textbf{F\textsuperscript{i}} & \textbf{F\textsuperscript{e}} & \textbf{F\textsuperscript{s}} \\ \midrule
\multirow{12}{*}{\rotatebox{90}{CoNSeP}}                                     
& Hover-Net     & 54.95 & 47.03 & 52.68 & 45.51 & 75.10 & 62.25 & 24.01 & 55.25 & 62.64 & 53.09 \\
& Hover-Net'    & 55.12 & 47.22 & 52.81 & 45.82 & 75.37 & 62.74 & 23.36 & 57.63 & 62.09 & 53.11 \\ \cline{2-12}
& SDM           & 55.15 & 47.54 & \textcolor{blue}{52.74} & 46.01 & 74.80 & 62.55 & 30.27 & 55.43 & 62.24 & 53.04    \\ 
& SDM+random    & 55.19 & 47.57 & 52.54 & 45.81 & 75.23 & 63.03 & 33.18 & 57.71 & 62.30 & 52.55    \\
& SDM+NMM       & 55.22 & \textcolor{blue}{47.70} & 52.65 & \textcolor{blue}{46.10} & 75.35 & 63.76 & \textcolor{blue}{40.72} & 58.17 & 62.58 & \textcolor{blue}{53.80}    \\ 
& SDM+IIM       & \textcolor{blue}{55.29} & 47.42 & \textcolor{blue}{52.74} & 45.97 & \textcolor{blue}{75.95} & \textcolor{blue}{64.07} & 37.35 & \textcolor{blue}{59.63} & \textcolor{blue}{62.97} & 53.15   \\ 
& IM-Diffusion           &  \textcolor{red}{55.62} &  \textcolor{red}{47.97} &  \textcolor{red}{53.20} &  \textcolor{red}{47.06} &  \textcolor{red}{76.26} &  \textcolor{red}{65.05} &  \textcolor{red}{46.67} &  \textcolor{red}{59.42} &  \textcolor{red}{63.15} &  \textcolor{red}{55.16}    \\ \cline{2-12}

& PNS           & 54.40 & 48.64 & 52.00 & 45.97 & 78.93 & 66.17 & 48.20 & 76.35 & 73.69 & 66.14   \\ 
& PNS'          & 54.43 & 48.81 & 52.19 & 46.30 & 78.48 & 66.23 & 48.18 & 76.11 & 73.79 & 65.83   \\  \cline{2-12}
& SDM           & 54.67 & 48.92 & 52.44 & 46.70 & 78.89 & 66.61 & 49.45 & 76.63 & 73.70 & 65.22    \\ 
& SDM+random    & 54.53 & 48.61 & \textcolor{blue}{52.57} & 46.47 & 78.79 & 66.86 & 50.78 & 76.29 & \textcolor{blue}{74.77} & 65.58     \\
& SDM+NMM       & 54.61 & \textcolor{blue}{49.31} & 52.42 & \textcolor{blue}{46.87} & \textcolor{blue}{79.43} & \textcolor{blue}{67.49} & \textcolor{blue}{52.85} & 76.69 & 74.73 & 66.22    \\ 
& SDM+IIM       & \textcolor{blue}{54.72} & 48.76 & 52.41 & 46.48 & 79.19 & 66.77 & 50.45 & \textcolor{blue}{76.76} & 74.23 & \textcolor{blue}{66.65}    \\ 
& IM-Diffusion           & \textcolor{red}{55.33} & \textcolor{red}{49.65} & \textcolor{red}{53.02} & \textcolor{red}{47.38} & \textcolor{red}{79.91} & \textcolor{red}{68.52} & \textcolor{red}{56.65} & \textcolor{red}{77.04} & \textcolor{red}{75.73} & \textcolor{red}{67.21}    \\  \bottomrule 

\end{tabular}
\end{table}